\documentclass{nature}
\usepackage{setspace}
\usepackage{amsfonts}
\usepackage{amssymb}
\usepackage{amsmath}

\usepackage{float}
\usepackage{placeins}
\usepackage{graphicx}

\usepackage[export]{adjustbox}

\usepackage[bf,bf,bf]{caption}

\renewcommand{\figurename}{Figure}
\renewcommand{\tablename}{Table}

\usepackage[super,comma,sort&compress]{natbib}

\usepackage{hyperref}
\usepackage{verbatim}

\usepackage{ulem} %provides various types of underlining that can stretch between words and be broken across lines
\usepackage{rotating} % rotate any object by arbitrary angle
\usepackage{soul}%h y p h e n -a t a b l e l e t t e r s p a c i n g ( s p a c i n g o u t ) , underlining and some derivatives such as overstriking and highlighting

\usepackage{color} %change color of text

\usepackage[colorinlistoftodos]{todonotes}

\makeatletter
\newsavebox\myboxA
\newsavebox\myboxB
\newlength\mylenA

\newcommand*\xoverline[2][0.75]{%
    \sbox{\myboxA}{$\m@th#2$}%
    \setbox\myboxB\null% Phantom box
    \ht\myboxB=\ht\myboxA%
    \dp\myboxB=\dp\myboxA%
    \wd\myboxB=#1\wd\myboxA% Scale phantom
    \sbox\myboxB{$\m@th\overline{\copy\myboxB}$}%  Overlined phantom
    \setlength\mylenA{\the\wd\myboxA}%   calc width diff
    \addtolength\mylenA{-\the\wd\myboxB}%
    \ifdim\wd\myboxB<\wd\myboxA%
       \rlap{\hskip 0.5\mylenA\usebox\myboxB}{\usebox\myboxA}%
    \else
        \hskip -0.5\mylenA\rlap{\usebox\myboxA}{\hskip 0.5\mylenA\usebox\myboxB}%
    \fi}
\makeatother

\usepackage{lscape}
\usepackage{mathtools}
\usepackage{gensymb}
\usepackage{graphicx}

\def\lapp{\ifmmode\stackrel{<}{_{\sim}}\else$\stackrel{<}{_{\sim}}$\fi}

\title{The first evidence for three-dimensional spin-velocity alignment in pulsars}

\author{
Jumei Yao$^{\rm 1,2}$$^\ast$,
Weiwei Zhu$^{1}$$^\ast$,
Richard N. Manchester$^{3}$,
William A. Coles$^{4}$,
Di Li$^{1,5}$$^\ast$,
Na Wang$^{2}$,
Michael Kramer$^{6,7}$,
Daniel R. Stinebring$^{8}$,
Yi Feng$^{1}$,
Wenming Yan$^{2}$,
Chenchen Miao$^{1}$,
Mao Yuan$^{1}$,
Pei Wang$^{1}$,
Jiguang Lu$^{1}$
}

\begin{document}
\maketitle
\begin{affiliations}
\item National Astronomical Observatories, Chinese Academy of Sciences, Chaoyang District, Datun Road, A.20, Beijing 100101, China 
\item Xinjiang Astronomical Observatory, Chinese Academy of Sciences, 150, Science 1-Street, Urumqi, Xinjiang 830011, China
\item CSIRO Astronomy and Space Science, Australia Telescope National Facility, P.O. Box 76, Epping NSW 1710, Australia
\item Electrical and Computer Engineering, University of California, San Diego, CA 92093, USA
\item University of Chinese Academy of Sciences, Beijing 100049, China
\item Max-Planck-Institut f\"ur Radioastronomie, Auf dem H\"ugel, 69, D-53121 Bonn, Germany 
\item Jodrell Bank Centre for Astrophysics, The University of Manchester, M13 9PL, UK
\item Department of Physics and Astronomy Oberlin College, Oberlin, OH 44074, USA\\
$^\ast$E-mail:yaojumei@xao.ac.cn, zhuww@nao.cas.cn, dili@nao.cas.cn
\end{affiliations}

{\bf More than 50 years after the discovery of pulsars \cite{hbp68} and confirmation of their association with supernova explosions \cite{lvm68,ccl+69,gol68}, the origin of the initial spin and velocity of pulsars remains largely a mystery. The typical space velocities of several hundred km~s$^{-1}$ have been attributed to ``kicks'' resulting from asymmetries either in the supernova ejecta or in the neutrino emission \cite{ll94,jan17,jb20}. Observations have shown a strong tendency for alignment of the pulsar space velocity and spin axis in young pulsars but, up to now, these comparisons have been restricted to two dimensions. We report here the first evidence for three-dimensional alignment between the spin and velocity vectors, largely based on observations made with the Five-hundred-meter Aperture Spherical radio Telescope  of the pulsar PSR~J0538+2817 and its associated supernova remnant S147. Analysis of these and related observations has enabled us  to determine the location of the pulsar within the supernova remnant and hence its radial velocity. Current simulations of supernova explosions have difficulty  producing such three-dimensional alignment \cite{wjm13,mth+19,jb20}. Our results, which depend on the unprecedented sensitivity of the new observations, add another dimension to the intriguing correlation between pulsar spin-axis and birth-kick directions, thus deepening the mysteries surrounding the birth of neutron stars.}

PSR J0538+2817 was discovered at Arecibo, and its short period (143 ms) and sky location within the boundary of SNR S147 immediately suggested that it was a young pulsar born in the supernova explosion that created S147 \cite{acj+96}. Modelling its surrounding X-ray torus structure, with the assumption of alignment between the torus's symmetry axis and the pulsar's spin axis, provided a position angle (PA) (i.e., the angle on the sky plane, measured from north toward east) of $\psi_X=154^\circ.0\pm5^\circ.5$ and an inclination angle from the line of sight of $\zeta_X=100^\circ\pm6^\circ$ into the plane of the sky \cite{rn03}. Early mean pulse polarization measurements of PSR J0538+2817 gave a value for $\zeta_{\rm pol}$ of $97^\circ$ with large uncertainties \cite{klh+03}; no measurement of $\psi_{\rm pol}$ was reported. Analysis of Very Long Baseline Array observations of PSR J0538+2817 gave an estimated distance of $D=1330\pm$190~pc and proper motions, referred to the local standard of rest (LSR), of $\mu_{\alpha}=-24.4\pm$0.10~mas~yr$^{-1}$ and $\mu_{\delta}=57.2\pm$0.10~mas~yr$^{-1}$ \cite{cbv+09, dny+15}. These correspond to a PA $\psi_{\rm pm}=337^\circ.0\pm0^\circ.1$, only $3^\circ\pm6^\circ$ from $\psi_X$ ($+180^\circ$), suggesting two-dimensional (2D) alignment. Radial velocities can in principle be determined from pulsar timing observations \cite{ehm06},  although no significant determination has yet been achieved. Optical lines from the binary companion \cite{fbw+11}, can be used to determine the radial velocity, but the pulsar velocity vector in such a system is essentially unrelated to any natal neutron-star kick.

Through the `shared-risk' early science program, we observed PSR J0538+2817 on four epochs between 2019 June and 2019 October with the Five-hundred-meter Aperture Spherical radio Telescope (FAST) \cite{fast18}, using the central beam of the 19-beam receiver covering the frequency band 1050~MHz to 1450~MHz. Here we focus on the October 11 (MJD 58767) observations as they had polarization calibration. For subsequent analyses, we extracted two bands, 1050–1150 MHz and 1350–1450 MHz, to avoid known radio frequency interference.

\begin{figure*}
\centering
\includegraphics[width=12.0 cm]{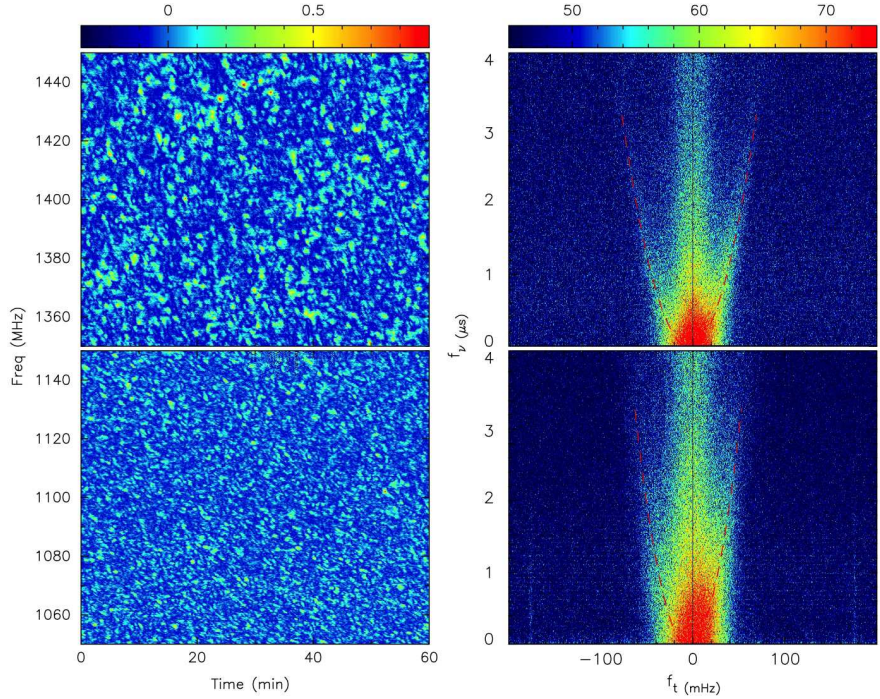}
    \caption{Dynamic spectra and secondary spectra for a 60-min FAST observation of PSR J0538+2817 for 100-MHz bands centred at 1100~MHz (lower panels) and 1400~MHz (upper panels). For the dynamic spectra the color scale is linear in signal power with arbitrary units. The secondary spectra are shown with a logarithmic (dB) color scale. The horizontal and vertical axes are conjugate time ($f_t$) and conjugate frequency ($f_\nu$). The red dashed lines represent the best-fit parameters of the arc, with the derived $\eta$ and $\gamma$ values scaled by $(1400/1100)^2$ and $(1400/1100)^3$ respectively for the 1100~MHz plot.}
    \label{fig:ds_ss}
\end{figure*}

The extremely small angular size of the pulsar emission region, coupled with multi-path propagation through an inhomogeneous interstellar electron distribution, produces a complex interference pattern at the observer \cite{ric90}. Since the interference phases are frequency dependent and there is relative motion between the pulsar, the scattering region and the Earth, we observe a ``dynamic spectrum" with interference maxima or ``scintles" as shown in the left panel of Fig.~\ref{fig:ds_ss}. The frequency and time scales of these scintles are strongly frequency dependent. Analysis of auto-correlation functions and structure functions show that the scintillation is strong and that the short refractive timescale is consistent with daily variations of the turbulence spectrum (see Methods).

The right panels of Fig.~\ref{fig:ds_ss} show the 2D power spectrum of each dynamic spectrum, known as the ``secondary spectrum". Both secondary spectra show an arc with a relatively sharp outer boundary and a central bright ridge along the $f_\nu$ axis at $f_t=0$. These scintillation arcs result from interference between the core of the brightness distribution and more highly scattered rays and require dominant scattering in a relatively thin screen along the path \cite{smc+01}. At both frequencies, the arcs show a clear asymmetry about the $f_t=0$ axis, indicating a persistent phase gradient across the scattering screen \cite{crs+06}. Fitting the more clearly delineated arc at 1400~MHz with $f_\nu=\eta f^2_t+\gamma f_t$ results in an arc curvature $\eta=(6.09\pm0.10)\times10^{-4}~\rm s^3$ and gradient parameter $\gamma=(4.84\pm0.30)\times10^{-6}~\rm s^2$,  corresponding to a ray deflection of $\theta_g=12.1\pm0.8$~$\mu$as across the scattering disk (see Methods for a full description of the procedure). Given the estimated distance to the pulsar and SNR, the derived pulsar-screen distance is $D_{\rm ps}=32.1\pm7.1$~pc, which is consistent with the radius $R_s=32.1\pm4.8$~pc of SNR S147, strongly suggesting that the scattering screen is located on the near side of the SNR shell (see Methods). 

As \ref{tb:obs_epochs} in Methods shows, $\eta$ is stable within the uncertainties across the four observational epochs,  but $\gamma$ varies considerably across intervals as short as one day. These results will be discussed in a future paper. Based on simulations, we attribute the rare central ridges observed for PSR J0538+2817 to scattering by asymmetric structures in the screen (see Methods for details).

\begin{figure*}
\center
\includegraphics[width=10 cm, angle=270]{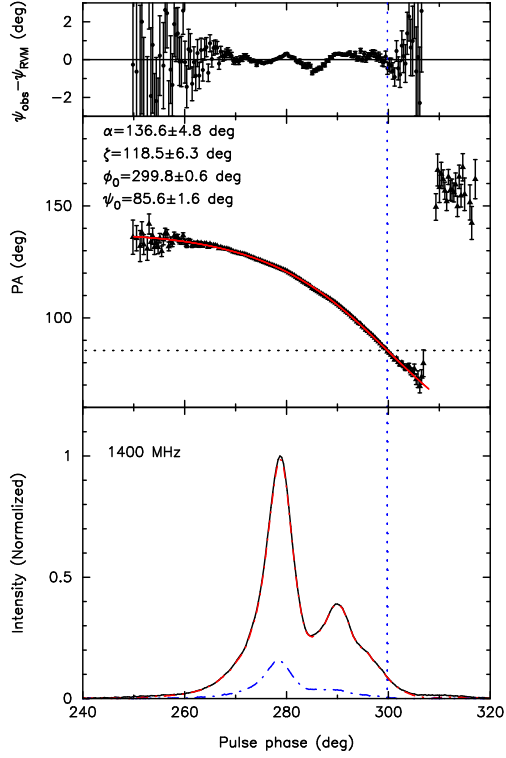}
    \caption{Polarization profile of J0538+2817 at 1400~MHz. The bottom panel shows the total intensity (solid black line), the linear polarization (red dash-dot line), and circular polarization (blue dash-dot line). The middle panel shows the observed position angles $\psi$ at 1400~MHz as a function of pulse phase and the red line gives the best-fit RVM solution. The top panel shows the fit residuals. The vertical dotted line shows the central pulse phase of the RVM fit, $\phi_0$, and the horizontal dotted line gives the corresponding PA at 1400~MHz, $\psi_0$. Best-fit parameters from the RVM fit are given in the middle panel.}
    \label{fig:poln_rvm}
\end{figure*}

In Fig.~\ref{fig:poln_rvm}, we show the polarization profile of PSR J0538+2817 derived from the FAST observations in the band 1350-1450~MHz. We derived the best-fit pulsar rotation measure (RM) by fitting the observed PAs across both observed FAST bands, 1050-1150~MHz and 1350-1450~MHz to give $RM=+39.56\pm0.14$~rad~m$^{-2}$. The red line in the middle panel of Fig.~\ref{fig:poln_rvm} represents the best-fit rotating-vector model (RVM) model \cite{rc69} to the observed PA variations (see Methods), and the top panel shows the corresponding fit residuals as a function of pulse phase. To compare $\psi_{\rm pol}$ with other determinations of the PAs of the projected spin axis and pulsar velocity, we must correct it to infinite frequency to give the so-called ``intrinsic” PA of the spin axis. Using the RM given above, we obtain $\psi_{\rm pol} (\rm intrinsic)=-18^\circ.5\pm1^\circ.7$. The polarization fitting gives an inclination angle of the pulsar spin axis to the line of sight of $\zeta_{\rm pol}=118^\circ.5\pm6^\circ.3$. 

\begin{figure*}
\center
\includegraphics[width=12.0 cm, angle=0]{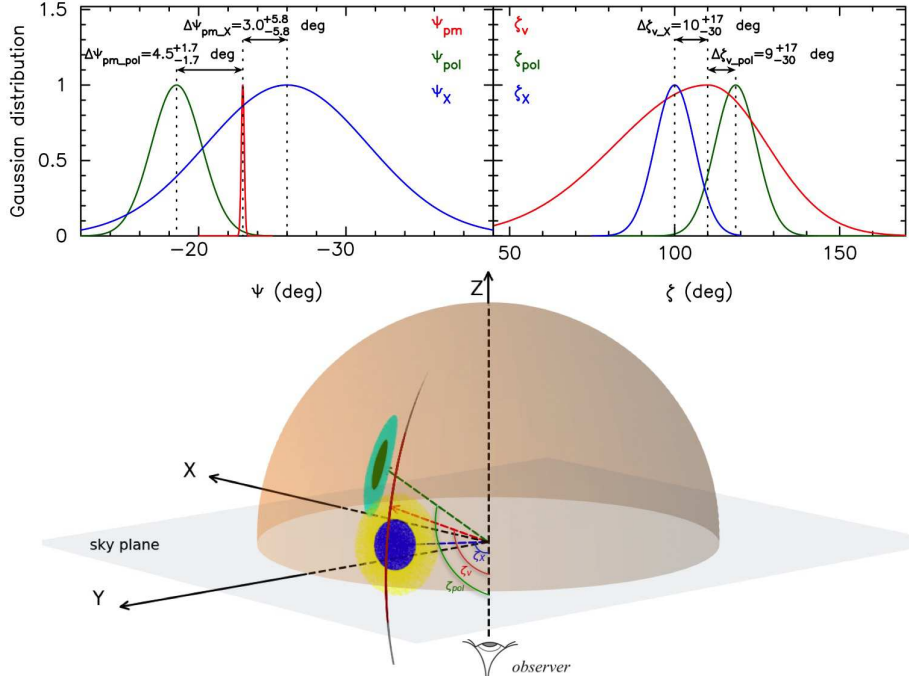}
    \caption{Position angles $\psi$ and inclination angles $\zeta$ of the spin axis and velocity vector (upper panels) and the corresponding 3D distribution on the surface of a unit sphere (lower panel). In the upper panels, the spin-vector constraints estimated from polarization PA fitting and X-ray torus fitting are shown in dark green and blue respectively, and the velocity-vector constraints obtained from VLBI (position angle) and the ISS analysis (inclination angle) are shown in red. The lower panel shows the 3D constraints with 68\% and 95\% confidence limit for the vector orientations, in dark green and light green for the spin obtained from PA fitting and in blue and yellow for the spin given by X-ray torus fitting, and in red and gray for the velocity, with the inclination angles marked.}
    \label{fig:3D}
\end{figure*}

Fig.~\ref{fig:3D} illustrates the 3D relationship of the derived pulsar velocity and spin vectors. On the plane of the sky, the position angle offset between the projected velocity and spin axis estimated from polarization PA fitting, $\Delta\Psi_{\rm pm\_pol}=4^\circ.5\pm1^\circ.7$, is consistent with that from modelling of the X-ray torus, $\Delta\Psi_{\rm pm\_X}=3^\circ.0\pm5^\circ.8$, further confirming approximate 2D alignment. The inclination angles of the velocity and spin vectors are less well determined. With the assumption that the scattering is dominated by the SNR shell and that the shell is spherical, we find that the pulsar is located $2.9^{+5.6}_{-5.4}$~pc behind the plane of the sky. This corresponds to a pulsar radial velocity of $81^{+158}_{-150}$~km~s$^{-1}$ and an inclination angle of the pulsar velocity of $\zeta_v=110^\circ\,^{+16^\circ}_{-29^\circ}$. This inclination angle overlaps well within 68\% confidence intervals with both $\zeta_{\rm pol}$ measured from FAST polarization ($\Delta\zeta_{\rm v\_pol}=9^\circ\,^{+30^\circ}_{-17^\circ}$) and $\zeta_{\rm X}$ measured from the X-ray torus modelling ($\Delta\zeta_{\rm v\_X}=10^\circ\,^{+30^\circ}_{-17^\circ}$). In the lower panel of Fig.~\ref{fig:3D}, we show the 3D relationships of the vectors with 68\% and 95\% confidence limit in their orientation. As Fig.~\ref{fig:3D} illustrates, the inclination angles for the spin vector derived from the pulse polarisation measurements and the X-ray torus measurements differ at about the 2-$\sigma$ level. We do not consider this difference to be significant given the possibility of unrecognised systematic errors in either or both measurements.

\begin{figure*}
\center
\includegraphics[width=12.0 cm, angle=270]{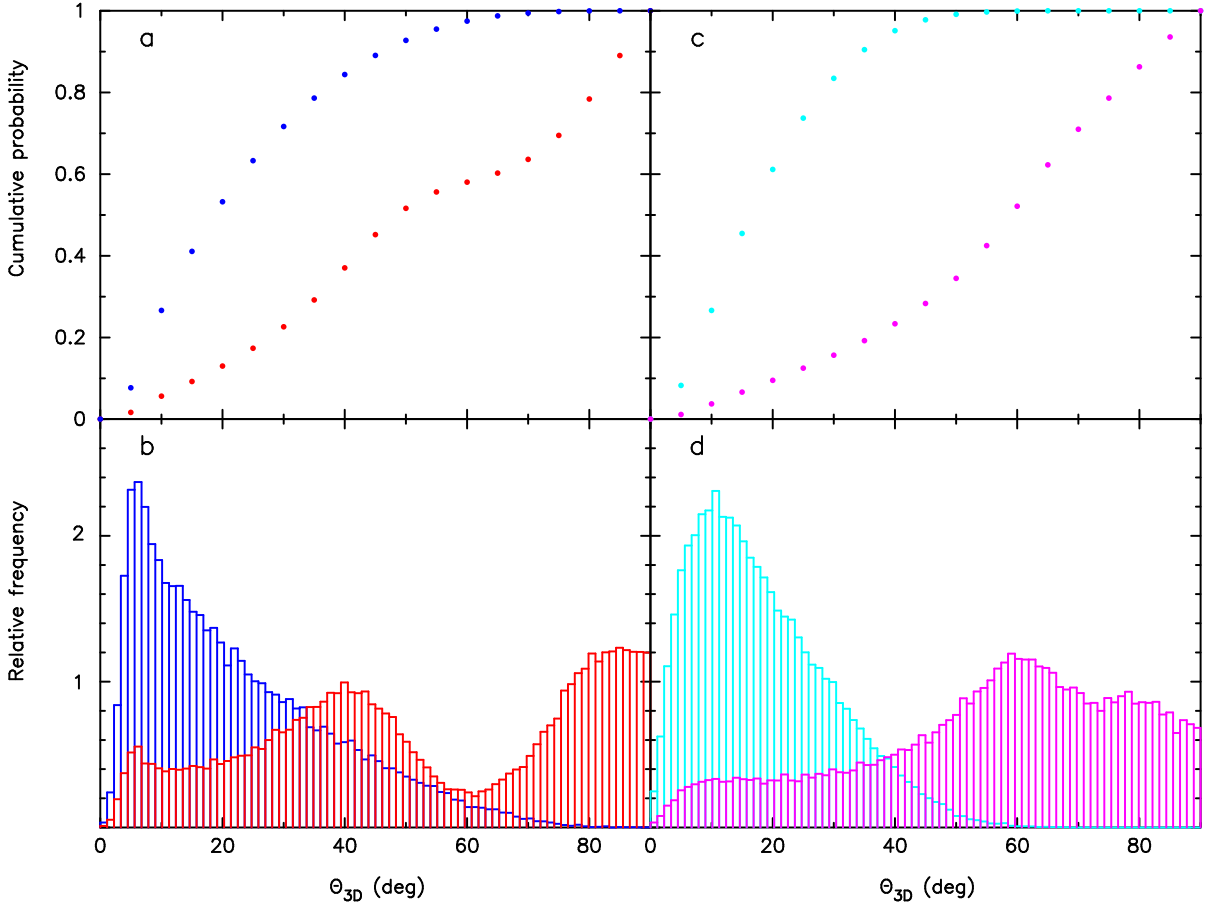}
    \caption{The cumulative and relative distributions of the spin-velocity misalignment angle $\Theta_{\rm 3D}$ that satisfy the constraints on $\psi$ and $\zeta$ shown in Fig.~\ref{fig:3D}, based on a random selection of 3D orientations for the spin, velocity and line-of-sight directions.  Spin vector constraints estimated from polarization PA fitting are shown in the left panel and from X-ray torus fitting in the right panel. For the red and magenta distributions, we ignore the observational constraints on the radial velocity and assume a uniform prior between $-$1500\;km\,s$^{-1}$ and +1500\;km\,s$^{-1}$. The blue and cyan distributions include all constraints shown in Fig.~\ref{fig:3D}}
    \label{fig:theta_3D}
\end{figure*}

Approximate 2D spin-velocity alignment has been demonstrated for more than 30 mostly young pulsars \cite{nr04,nkc+12}. However, this is the first time that 3D alignment of a pulsar's spin and space-velocity vectors has been demonstrated. The significance of this result is illustrated in Fig.~\ref{fig:theta_3D}, which compares the constraints on the relative alignment of the spin and velocity vectors with and without the knowledge of the measured pulsar radial velocity.
For the latter, following Din{\c{c}}el et al.\cite{dny+15}, we assume a uniform prior between $-$1500\;km\,s$^{-1}$ and +1500\;km\,s$^{-1}$. With this assumed radial velocity distribution, the distribution of $\Theta_{\rm 3D}$ based on the FAST polarization results (shown in red in Fig.~\ref{fig:theta_3D}) has a large 68\% confidence interval of about 74$^\circ$. For the X-ray torus results, the distribution (shown in magenta in Fig.~\ref{fig:theta_3D}) is also broad with a 68\% confidence interval of about 68$^\circ$. Including the observed constraint on $\zeta_v$ strongly constrains the 3D angle between the spin axis measured from FAST polarization and the velocity vector with a clear peak around 6$^\circ$ and a 68\% probability of being less than $28^\circ$. From the X-ray torus modeling there is a peak around 10$^\circ$ and the angle is smaller than $23^\circ$ with 68\% probability. It is clear that the observational constraints strongly limit the range of possible misalignment angles. For the polarization results, the low probability of misalignment angles less than $<$4$^\circ$ results from the tight constraint on the position angle of the transverse velocity from the VLBI proper-motion results (see Fig.~\ref{fig:3D}, upper left panel).

Existing models struggle to explain the spin-velocity 3D alignment that we have observed for PSR J0538+2817 and which, for other pulsars, is commonly assumed based on the 2D alignment. For example, the electromagnetic-rocket model \cite{th75} invokes a post-natal acceleration along the pulsar's spin axis by an asymmetric radiation-reaction force from an offset magnetic dipole.  This model does result in spin-velocity alignment, but requires a very high initial neutron-star spin rate ($>$1\;kHz), which is unlikely for PSR J0538+2817, especially given its long inferred  initial spin period of 140\;ms \cite{klh+03}. Numerical modelling of supernova explosions in order to determine the explosion energetics and the properties of any compact remnant has been carried out for decades. Initially such modelling was in 1D, then 2D and more recently in 3D, with a wide variety of numerical codes emphasizing different aspects of the explosion and making different approximations \cite{jms16}. In general, 3D models do not include rotation of the progenitor star, since rapidly spinning iron cores are not expected for the majority of supernova progenitors \cite{hws05}. In these simulations, neutron-star kicks could result from asymmetries in the ejecta (e.g., the ``tug-boat" mechanism) or in the neutrino emission, and spins result from fall-back of ejecta onto the neutron star \cite{jan17, jb20}. In these models, there is little or no relationship between the neutron-star kick direction and spin axis and hence no tendency for an alignment. Possible mechanisms for spin-velocity alignment include ejecta directed by the rotation of the progenitor star \cite{jan17,mth+19,jb20}, neutrino emission preferentially channelled along the rotation axis \cite{nr07} and random torques averaged over several rotation periods of the proto-neutron star producing a net ``slow kick" along the spin axis \cite{sp98,nr07}, but there is no generally accepted explanation. 

It is also clear that the alignment is not perfect. Based on existing 2D spin-velocity alignment detections, small but significant misalignments of $5^\circ$ - $10^\circ$ have been observed in many pulsars, including PSR J0538+2817. These misalignments could result from break-up of a pre-supernova binary system \cite{dny+15} or from stochastic variations in the direction of the natal kick or spin-up torques \cite{jan17,jb20} and are an important diagnostic of the processes involved. Systematic measurements of the 3D orientations of pulsar spin and velocity vectors, building on the present work, are thus highly desirable. With its great sensitivity, FAST is poised to play an important role in these investigations.

\clearpage

\pagestyle{empty}
\renewcommand{\thesubsection}{\Alph{subsection}}
\renewcommand{\thefigure}{\thesubsection .\arabic{figure}}
\renewcommand{\thetable}{\thesubsection .\arabic{table}}
\setcounter{figure}{0}
\setcounter{table}{0}

\begin{methods}
\subsection{Observations and data processing.}
The polarization response was calibrated using short observations of a pulsed noise source injected into the feed before and after the pulsar observation. Since the observations were at low zenith angle, the system temperature $T_{\rm sys}\leq 20$~K which, combined with the system gain of $\sim 16$~K/Jy \cite{jth+20}, enabled a single pulse detection threshold of 0.25 mJy (5$\sigma$). A digital filterbank system was used to form and record the full polarization data sampled at 49.152~$\mu s$ intervals with 3277 channels of bandwidth of 0.122~MHz across the analyzed 400-MHz band from 1050~MHz to 1450~MHz. We used the analysis program {\sc dspsr} \cite{sb11} and the {\sc psrchive} software package \cite{sdo12} to reduce our data. Data for each channel were folded according to the pulsar period using a sub-integration time of 2.5~s and then polarization calibrated.

\subsection{Interstellar scintillation data analysis.}
For the ISS analysis, we first formed total intensity (Stokes $I$) profiles for each channel and sub-integration, did further RFI rejection using {\sc pazi}, and then used {\sc psrflux} to form the dynamic spectrum. To remove the effects of short-term pulse intensity variations, we normalize the mean power of each sub-integration spectrum to the mean power across the observation. To compute the secondary spectra, we applied a Hamming window function to the outer 10\% of each dynamic spectrum to reduce the effects of aliasing in the secondary spectrum. Since the scintillation fluctuation spectrum is typically steep, we use first-differences to pre-whiten the dynamic spectrum \cite{chc+11}. After this, we form the secondary spectrum using a 2D Fourier transform, taking its squared magnitude, and finally applying the appropriate post-darkening filter to recover our best estimate of the secondary spectrum. The Nyquist frequencies corresponding to the 2.5~s sub-integration time and the 0.122~MHz channel bandwidth are $f_t(\rm Nyquist) = 200$~mHz and $f_\nu(\rm Nyquist)=4.1$~$\mu$s, respectively.

\subsection{Polarization data analysis.}
For the polarization analysis, we summed the 60-min observation in time to form a single calibrated polarization profile for each channel. Using the {\sc psrchive} program {\sc rmfit}, we derived the best-fit pulsar RM by fitting the observed PAs across both observed FAST bands, 1050-1150~MHz and 1350-1450~MHz and obtain $RM=+39.56\pm0.14$~rad~m$^{-2}$. This is quite different to the previously published value: $RM=-7\pm12$~rad~m$^{-2}$ \cite{mwk+03}. We have no explanation for this difference, but we are confident that our measurement accurately gives the RM at the time of the FAST observations. In the RVM convention, the PA as a function of pulse longitude, $\phi$, can be expressed as

\begin{equation}
\psi=\psi_{\rm pol}+\rm arctan\displaystyle\frac{\sin\alpha\sin(\phi-\phi_0)}{\sin\zeta\cos\alpha-\cos\zeta\sin\alpha\cos(\phi-\phi_0)}
\end{equation}
where $\phi_0$ is the pulse phase for the closest approach of the line of sight to the magnetic axis, with a corresponding PA $\psi_{\rm pol}$. In the simple dipolar model, $\psi_{\rm pol}$ is the position angle of the pulsar spin axis, $\alpha$ is the angle between the spin axis and magnetic axis, and $\zeta$ is the inclination angle of the spin axis from the line of sight. Choosing data from the 1350-1450 MHz band, we then fitted the RVM model to the PA variation across the profile using {\sc psrmodel}; the resulting fit is shown in Fig.~\ref{fig:poln_rvm}. The fit residuals shown in the upper panel are very close to zero across the two main pulse components (where the S/N is sufficient). However, there are significant deviations in the region where the two components overlap, suggesting that the two components come from (slightly) different source regions. Mixing of the radiation from the two regions could result in the observed small deviations in the resultant PA in the overlap region.

In order to compare $\psi_{\rm pol}$ with other determinations of the PAs of the projected spin axis and pulsar velocity, we must correct it to infinite frequency to give the so-called ``intrinsic” PA of the spin axis. Using the RM given above, we obtain $\psi_{\rm pol} (\rm intrinsic)=-18^\circ.5\pm1^\circ.7$. The inclination angle of the pulsar spin axis to the line of sight needs no correction and is $\zeta_{\rm pol}=118^\circ.5\pm6^\circ$.3. The derived inclination angle of the magnetic axis relative to the spin axis, $\alpha = 136^\circ.6\pm4^\circ.8$, implying a minimum angle between the magnetic axis and the line of sight $\beta=\alpha-\zeta=18^\circ.1\pm7^\circ.9$ with an ``outer” line of sight, that is, passing on the equatorial side of the magnetic axis \cite{lm88}. Note that our analysis uses the IAU convention of PA increasing in a counter-clockwise direction on the sky. {\sc psrmodel} uses the opposite sense for PA and we have used conversion equations to obtain a consistent set of parameters \cite{ew01}.

\subsection{ACF and structure function analysis.}
We form two-dimensional auto-correlation functions (ACFs) of the intensity fluctuations, $\rho_{\rm int}(\tau,\nu)$, and then estimate the diffractive time and frequency scales, $\Delta\tau_d$ and $\Delta\nu_d$ respectively. As $\rho_{\rm int}(\tau,\nu)$ shows a skewness resulting from a phase gradient across the scattering disk, the frequency widths are somewhat biased. Hence, we can only determine approximate scintillation bandwidths $\Delta\nu_d$ from the 2D ACF $\rho_{\rm int}(\tau,\nu)$, giving values of 0.40~MHz and 0.86~MHz at 1100~MHz and 1400~MHz respectively. The scintillation strength $u$, defined as $u = \sqrt{\nu/\Delta\nu_d}$ \cite{ric90}, is 52 and 40 respectively, indicating that the scintillation is very strong in both observational bands. 

A cut through the temporal axis of the ACF, $\rho_{\rm int}(\tau, 0)$, is independent of any phase gradient and hence can be used to estimate the phase structure function D$_{\phi}(\tau)$ and the the power spectrum of the turbulence. The canonical problem in small-angle forward scattering from a thin phase screen is appropriate for cases of interstellar scintillation in which the scattering plasma occupies a small fraction of the line of sight. We consider a plane wave in free space propagating in the z direction incident on a thin slab of material of thickness $\delta z$ with a refractive index $n(x,y,0) \ne 1$. On passing through this slab the electric field is $E(x,y,0) = \exp(-j\phi(x,y))$, where $\phi(x,y) = k n(x,y,0)\delta z$ and $k = 2 \pi / \lambda$. This field can be expressed as an angular spectrum of plane waves by taking the 2D Fourier transform of E(x,y).

\begin{equation}
    \tilde{E}(\kappa_x,\kappa_y) = \int\!\!\int E(x,y)\exp(-j(\kappa_x x + \kappa_y y)) dx\,dy
\end{equation}
Here the plane waves are described by $\kappa_x = k \sin( \theta_x)$ and $\kappa_y = k \sin(\theta_y)$. The ACF of the electric field, which is called the ``mutual coherence function'' in statistical optics, is given by

\begin{equation}
    \rho_e(s_x,s_y,z=0) = \langle \exp(-j(\phi(x,y)-\phi(x+s_x,y+s_y))) \rangle.
\end{equation}
If $\phi(x,y)$ is a Gaussian random process then 

\begin{equation}
\rho_e(s_x,s_y,z=0) =  \exp(-0.5 \langle(\phi(x,y)-\phi(x+s_x,y+s_y))^2 \rangle ).
\end{equation}
The argument of the exponential is called the phase structure function $D_\phi (s_x, s_y)$.  If the power spectrum of $\phi(x)$ is power-law $\sim \kappa_x^{-\alpha}$ with $\alpha < 4$ then the the structure function is power-law $\sim s_x^{\alpha-2}$. Thus a measurement of $D_\phi (s_x,s_y)$ provides an estimate of the spectral exponent (for which the ``Kolmogorov" value = 11/3). The Fourier transform of $\rho_e(s_x,s_y)$ is the angular power spectrum $B(\theta_x,\theta_y)$. The plane wave changes phase as it propagates to the observer but it does not change amplitude. So $B(\theta_x,\theta_y)$ does not change as the angular spectrum propagates from the slab to the Earth, and $\rho_e(s_x,s_y)$ cannot change either. Thus $\rho_e(s_x,s_y)$ could be measured at the Earth and it would still be defined by $D_\phi (s_x,s_y)$.

In general one would need an interferometer to measure $\rho_e(s_x,s_y)$ at the Earth (where it would be called the ``visibility"). However as the wave propagate from the slab to the observer, the fluctuations in E(x,y) develop amplitude fluctuations. When these are ``strong" enough, E(x,y) becomes a 2D Gaussian random variable. It is much easier to measure the intensity than the complex electric field and in this case we have
$\rho_{\rm int}(s_x,s_y)= |\rho_e(s_x,s_y)|^2$.  Normally we measure intensity $I(t) = I(x = Vt)$ so $\rho_{\rm int}(\tau) = |\rho_e (V\tau,0)|^2$. Hence, we obtain the expression 

\begin{equation}
    D_{\phi}(\tau)=-\rm log_e[\rho_{\rm int}(\tau)]
\end{equation}
to derive the phase structure function $D_{\phi}(\tau)$ from the 1D temporal intensity ACF $\rho_{\rm int}(\tau)$.\footnote{The structure function is a useful statistic for other applications, particularly when the data series is short compared with the time scale of the fluctuations in it. It is also very useful for power-law processes which do not have a finite correlation function. So one must be careful to specify exactly what variable the structure function is describing. The structure function of phase is sometimes called the ``wave structure function'' in optical propagation.}

 To accurately determine $\rho_{\rm int}(\tau)$ for each band, we divide the dynamic spectra into 10 sub-bands, each 10~MHz wide, form the time-domain ACF for each, and then average these to get the mean ACF for each band. The zero-lag (or total power) point on these ACFs includes a white noise contribution, mainly from receiver noise. This white noise contribution was estimated by removing power from the zero-lag component until the peak of the ACF was approximately Gaussian and then renormalising the ACF. After this correction, $D_{\phi}(\tau)$ was approximately linear (on log-log axes) at small $\tau$.

In Extended Data\ref{fig:acf_sf} the upper panels show 1D ACFs for the two bands for observations on two consecutive days, 2019 October 10 and 11. The white noise spikes in the ACFs before correction were: for MJD 58766, 11.0\% and 9.1\% of the total power at 1100 and 1400~MHz, respectively, and for MJD 58767, 4.8\% and 6.1\% of the total power. Using a Gaussian function to fit the peak of the ACFs for MJD 58767, we find that the scintillation timescales $\Delta \tau_d$ at 1100 and 1400~MHz are 19.1$\pm$0.3 and 29.1$\pm$1.1~s, respectively. Here the uncertainties include two parts, the statistical uncertainty from the data fitting and the fractional uncertainty from the finite number of observed scintles in the dynamic spectra \cite{cwb85, wmj+05}.

Phase structure functions corresponding to these ACFs are shown in the lower panels of Extended Data\ref{fig:acf_sf}. Here we have scaled $D_{\phi}(\tau)$ by (1100/1400)$^2$ for the 1100~MHz band, to make the comparison with $D_{\phi}(\tau)$ at 1400~MHz exact. For MJD 58767, the fitted slopes for both bands are broadly consistent with the expected Kolmogorov exponent of 5/3 for $\tau < 20$~s, but there is a suggestion of flattening at longer time lags in the 1400~MHz data, which is more reliable at longer time lags. This flattening is much more prominent in the observations on the previous day MJD 58766, as shown in lower left panel of Extended Data\ref{fig:acf_sf}, and the fitted slopes of $\sim$1.2 are inconsistent with a Kolmogorov fluctuation spectrum. The time scale for the line of sight to cross the scattering disc (the refractive timescale) is $\Delta \tau_r=u^2\Delta \tau_d$ and so about 14.3~h at 1100~MHz and 12.9~h at 1400~MHz. Consequently, daily variations of the turbulence spectrum are consistent with the estimated refractive timescale.

\subsection{Arc curvature fitting.}
Scintillation arcs may be described by the equation

\begin{equation}
   f_\nu = \eta f_t^2 + \gamma f_t 
\end{equation}
where $\eta$ is the arc curvature and $\gamma$ represents a phase slope across the scattering disk \cite{crs+06}. To estimate $\eta$ and $\gamma$ we use the more clearly delineated asymmetric arc at 1400~MHz. We first randomly generate 5000 pairs of ($\eta$, $\gamma$) over the ranges $\eta = 4.6\times10^{-4}$ to $7.6\times10^{-4}$~s$^3$ and $\gamma=-5\times10^{-6}$ to $15\times10^{-6}$~s$^2$. Each ($\eta$, $\gamma$) corresponds to an arc in the secondary spectrum. We then sample the outer parts of each arc ($f_t\in[-80~\rm mHz, -40~\rm mHz]~~and~~[40~\rm mHz, 80~\rm mHz]$) with 350 points on each side, taking the nearest point in the secondary spectrum to each point along the arc, weighting it by its normalised power, and forming the summed weighted power for that arc.  We then use a 2D Gaussian function to fit the relatively symmetric peak of the arc power distribution over the ranges $\eta = 5.4\times10^{-4}$ to $6.8\times10^{-4}$~s$^3$ and $\gamma=1\times10^{-6}$ to $9\times10^{-6}$~s$^2$ (see Extended Data\ref{fig:arc_fit}) resulting the following best estimates for $\eta$ and $\gamma$:

\begin{equation}
    \eta=(6.09\pm0.10)\times10^{-4}~~~\rm s^3
\end{equation}
and

\begin{equation}
    \gamma=(4.84\pm0.30)\times10^{-6}~~~\rm s^2.
\end{equation}
The quoted uncertainties are calculated by dividing the 60~min observation into $4\times15$-min blocks, separately fitting for $\eta$ and $\gamma$ using the procedure described above for each block and taking one half of the rms deviation about the mean as the uncertainty in each parameter. 

For $f_\nu$ and $f_t$, we have

\begin{equation}
    f_\nu=\displaystyle{\frac{D(1-s)}{2cs}}(\theta^2+2\theta\theta_g)
\end{equation}
and

\begin{equation}
    f_t=\displaystyle\frac{2\pi\nu_c V_{\rm eff,\perp}}{cs}\theta,
\end{equation}
where $\nu_c$ is the band-centre frequency, $\theta$ is the ray deflection angle, $\theta_g$ is the refractive angle of the wave-front, and the scattering screen is located at a distance $sD$ from the pulsar \cite{crs+06}. The effective perpendicular (or transverse) velocity is given by

\begin{equation}\label{eq:v_eff}
    V_{\rm eff,\perp} = (1-s) V_{\rm psr, \perp}  + s V_{{\rm Earth},\perp} - V_{{\rm scr},\perp}
\end{equation}
where $V_{\rm psr, \perp}$ is the pulsar transverse velocity obtained from the measured proper motion, $V_{{\rm Earth},\perp}$ is the component of the Earth's velocity perpendicular to the line of sight to the pulsar, and $V_{{\rm scr},\perp}$ is the component of the velocity of the scattering screen perpendicular to the line of sight. After correcting the pulsar proper motion to the LSR \cite{dny+15} and taking the pulsar distance of 1330~pc, the pulsar perpendicular velocity ($V_{\rm psr, \perp}$) is $391\pm56$~km\,s$^{-1}$. Even with correction to the LSR, the Earth's space velocity is $\lapp 30$~km\,s$^{-1}$. We can therefore safely neglect the Earth velocity in Equation~\ref{eq:v_eff} (especially since, as we show below, $s$ is small). Although we are not sure of the exact location or identity of the scattering screen, there are an abundance of filaments located in the expanding gas behind the outer shock which are good candidates for the scattering screen. These filaments have an expansion velocity of about 80~km\,s$^{-1}$ with a scatter of order 20~km\,s$^{-1}$ \cite{ka79}, and the perpendicular component of this is $35\pm 9$~km\,s$^{-1}$.  Including this term, we have

\begin{equation}
    V_{\rm eff,\perp} \approx (1-s) V_{\rm psr, \perp} - V_{\rm scr, \perp};
\end{equation}
\begin{equation}\label{eq:eta}
    \eta=4625\displaystyle\frac{D_{\rm kpc}s(1-s)}{\nu^2_{\rm GHz}[(1-s)V_{\rm psr,\perp}-V_{\rm scr, \perp}]^2} 
\end{equation}
and

\begin{equation}\label{eq:gamma}
    \gamma=0.1496\displaystyle\frac{D_{\rm kpc}\theta_{\rm g, mas}(1-s)}{\nu_{\rm GHz} [(1-s)V_{\rm psr, \perp}-V_{\rm scr, \perp}]},
\end{equation}
where velocities are in km\. Taking $\nu=1.4$~GHz and assuming Gaussian distributions for the measured values of $V_{\rm psr, \perp}$, $V_{\rm scr, \perp}$, $\eta$, $\gamma$ and $D$ in Equations~\ref{eq:eta} and \ref{eq:gamma}, we obtain the distributions of $\theta_g$ and $s$ using a Monte Carlo analysis. Then we fit the distributions with Gaussian functions having different widths on the two sides and find that

\begin{equation}
    \theta_g=12.1\pm0.8~\rm \mu as,
\end{equation}
and

\begin{equation}
    s = 0.0241 \pm 0.0041.
\end{equation}
The uncertainty of $s$ is dominated by the uncertainty in the pulsar distance. This contributes 93\% of the total uncertainty, compared to 6\% from the uncertainty in screen velocity and 1\% from the uncertainty in the measured scintillation arc curvature and pulsar proper motion.

This determination of $s$ is based on observations made on MJD 58767. As mentioned in the main paper, scintillation observations of J0538+2817 were also made at three earlier epochs. The results from all observations are summarized in \ref{tb:obs_epochs}. This table shows that the arc curvature $\eta$ is relatively stable over the 132 days covered by these observations, with an rms scatter of $0.08\times 10^{-4}~{\rm s}^3$, consistent with the uncertainty of each measurement. This corresponds to a variation in $\zeta_v$ of less than 5$^\circ$. On the other hand, the arc asymmetry $\gamma$ and hence refractive angle $\theta_g$ vary greatly from epoch to epoch. These time variations and their implications will be discussed in a future publication.

Substituting for the measured values, the pulsar--screen distance $D_{\rm ps} = sD$ is $32.1\pm 7.1~{\rm pc}$. The uncertainty in $D_{\rm ps}$ is given by

\begin{equation}
    \delta D_{\rm ps}=\sqrt{D^2\;\delta s^2+s^2\;\delta D^2},
\end{equation}
where $\delta s$ and $\delta D$ are the uncertainties of $s$ and $D$, respectively. Similar to $s$ itself, the dominant contribution to  $\delta D_{\rm ps}$  is from the uncertainty in $D$ (96\%).

\subsection{The association between PSR J0538+2817 and SNR S147.}
S147 is an almost circular SNR of radius $\theta_s = 83^{\prime}\pm3^{\prime}$ \cite{shf80}, located in the Galactic anti-centre region with geometric centre at $\alpha_{\rm snr}=$05h~40m~01s$\pm$2~s, $\delta_{\rm snr}=+27^\circ\;48^{\prime}\;09^{\prime\prime}\pm20^{\prime\prime}$ \cite{klh+03}. PSR J0538+2817 is located at $\alpha_{\rm psr} =$~05h~38m~25$^{s}$.0572, $\delta_{\rm psr}=+28^\circ\;17^{\prime}\;09^{\prime\prime}.161$ \cite{cbv+09}. The angular offset of the pulsar from the SNR centre, $\theta_p$, is given by

\begin{equation}
    \theta^2_p = (\alpha_{\rm snr}-\alpha_{\rm psr})^2\,\cos^2\delta_{\rm snr} + (\delta_{\rm snr} - \delta_{\rm psr})^2,
\end{equation}
i.e., $\theta_p = 35^{\prime}.9 \pm 0^{\prime}.4$. Based on extinction in associated dust clouds, the estimated distance to S147 is $1220\pm 210$~pc \cite{clr+17}, consistent with the VLBA pulsar distance of $1330\pm 190$~pc \cite{cbv+09}, giving a physical radius for the SNR of $R_s = D\,\theta_s = 32.1\pm 4.8$~pc.

Within the uncertainties, this is equal to the estimated pulsar--scattering screen distance derived above, $D_{\rm ps} = 32.1 \pm 7.1$~pc, strongly suggesting that the scattering screen can be identified with the near side of the SNR shell and that the pulsar is within the shell. However, it is possible (although unlikely) that the pulsar could be behind the SNR and that the scattering screen causing the observed arc is the far side of the SNR shell. In this case, we might also expect to see scattering from the near side of the shell, about 100~pc from the pulsar. There is no evidence for such a second arc, supporting the idea that the pulsar is located within the shell and that the scintillation arc results from scattering in the near side of the SNR shell.

Overall, there is a low probability ($\sim 3\times10^{-4}$) of having a random young pulsar like PSR J0538+2817 pass so close to the centre of S147 and not be associated at birth \cite{nr04}. These arguments together with the strong evidence presented above that the pulsar lies within the SNR shell and is scattered by it make the case for the physical association of PSR J0538+2817 and SNR S147 overwhelming.

\subsection{The calculation of the inclination angle.}
To compute $\zeta_v$, the inclination angle of pulsar velocity vector from the observer, we adopt a coordinate system with the geometry showed in Extended Data\ref{fig:coord} with assumptions that SNR S147 is a perfectly spherical SNR, and that the scattering screen is located exactly at the outer edge of the SNR shell, i.e. $r = R_s$. From the right-triangle OKS, we have

\begin{equation}
    (D_{\rm ps}-z_p)^2+ x^2_p + y^2_p = R^2_s.
\end{equation}
Solving this for $z_p$ and neglecting the unphysical solution, we have

\begin{equation}
    z_p=D_{\rm ps}-\sqrt{R^2_s-x^2_p-y^2_p}= D\,\left ( s - \sqrt{\theta_s^2 - \theta_p^2} \right ) = D\,\theta_z,
\end{equation}
where we define the dimensionless scale factor $\theta_z$. Assuming Gaussian distributions for the input parameters, we get the distribution of $z_p$ using a Monte Carlo analysis. Following the same procedure used for estimating $\theta_g$ and $s$, we obtain

\begin{equation}
    z_p = 2.9^{+5.6}_{-5.4}\;\;{\rm pc}.
\end{equation}
 
Since we can assume that PSR J0538+2817 was born at the centre of SNR S147, we have $V_z=z_p/\tau_k$ where $\tau_k$ is the kinematic age. Given the improved proper motion measurements \cite{cbv+09}, we estimate a revised kinematic age of $\tau_k =34.8\pm 0.4$~kyr, and obtain the pulsar radial velocity

\begin{equation}
    V_z = 81^{+158}_{-150}\;\; {\rm km~s}^{-1}
\end{equation}
and a total space velocity

\begin{equation}
    V = 407^{+79}_{-57}\;\; {\rm km~s}^{-1}.
\end{equation}
From triangle OPS, using the Law of Cosines and defining $\Theta_p$, we obtain

\begin{equation}
   \zeta_v= 180^\circ - \rm arccos \left(\displaystyle\frac{D^2_{\rm op}+D^2_{\rm ps}-R^2_s}{2\,D_{\rm op}\,D_{\rm ps}}\right)= 180^\circ - \rm arccos\left (\displaystyle \frac{\Theta_p^2 +s^2 - \theta_s^2}{2\,s\,\Theta_p} \right ), 
\end{equation}
where

\begin{equation}
    D_{\rm op}^2 = x_p^2 + y_p^2 + z_p^2= D^2\,(\theta_p^2 + \theta_z^2) = D^2\,\Theta_p^2.
\end{equation}
Again following the procedure for estimating the uncertainty in $\theta_g$ and $s$, we obtain

\begin{equation}
    \zeta_v={110^\circ}^{+16^\circ}_{-29^\circ}
\end{equation}
for the inclination angle of the pulsar velocity vector.

We note that the uncertainty in $z_p$, and hence in $V_z$ and $\zeta_v$, is dominated by the uncertainty in the distance $D$ to the pulsar (and to the SNR). This contributes about 87\% of the uncertainty in $z_p$. The derived value of $z_p$ also depends on the assumptions that the scattering is dominated by a thin screen located at the boundary of a spherical SNR. As discussed in the previous section, filamentary structure in the shocked gas immediately behind the outer shock is a very plausible location for the scattering screen. The H$\alpha$ images of S147 \cite{ka79} suggest that most of the filaments lie close to the outer boundary of the shell, say within about 10\% of the boundary radius. The estimated fractional uncertainty in the shell radius, $\theta_s$, is about 4\% \cite{shf80}, although particular regions could have departures from sphericity larger than that. Added in quadrature, the uncertainties in $\theta_s$ and screen location relative to the shell add about 1.0~pc to the uncertainty in $z_p$, small compared to its estimated uncertainty of 5.5~pc. $V_z$ and $\zeta_v$ have additional uncertainties related to the uncertainty in the exact location of the pulsar birthplace, but their contributions are also small compared to that from the estimated distance of 87\%. While any improvement in the relevant parameters is desirable, improved VLBI astrometry for the pulsar and/or an improved distance estimate for the SNR would be the most valuable.

Regarding possible origins for the (small) misalignment of the spin and velocity vectors for PSR J0538+2817, it is of interest that {Din{\c{c}}el} et al. \cite{dny+15} have identified an OB runaway star within SNR S147. The OB star has a proper motion consistent with an origin at the estimated birth location of PSR J0538+2817 and a 3D velocity which is of opposite sign to the pulsar velocity, but with a substantial misalignment. Break-up of a pre-supernova binary system containing this star would have perturbed any pulsar velocity arising from a natal kick.

%As the measured total velocity of PSR J0538+2817 is smaller than 600 km\,s$^{-1}$, it's difficult to determine whether this pulsar has experienced the envelope stripping and received larger natal kick during its formation when compared with single stars \cite{spb20}. But break-up of a pre-supernova binary system containing this star would have perturbed any pulsar velocity arising from a natal kick.

\subsection{Modeling the central bright ridge}
The central bright ridge could be caused by instrumental effects, such as fine scale variations in the system bandpass which also varied with time on a scale of $\sim$30~s. However, no ridge is observed in any of several other pulsars observed with the same receiver on FAST and showing clear parabolic arcs. Alternatively, the ridge could represent a collection of high-curvature arcs caused by low-velocity scattering in the SNR or scattering by a screen nearer to midway along the line of sight. Neither of these is likely because the velocity is dominated by the pulsar proper motion and scintillation due to a screen along the path would be quenched by the apparent angular size of the pulsar caused by scattering in the SNR shell. Such an arc-ridge structure can also arise if the scattering medium is anisotropic with the transverse velocity aligned with the major axis of the scattering irregularities.

To explain this, we modeled the secondary spectra having arcs with sharp outer edges as a mapping from the brightness distribution B($\theta
_x, \theta_y$) to the secondary spectrum S($f_t , f_\nu$), where 

\begin{equation}
  f_t=\displaystyle\frac{\theta_x V_x}{\lambda} \;\;{\rm and}\;\; f_\nu = \displaystyle\frac{(\theta_x^2 + \theta_y^2) Z}{2c}
\end{equation}
\begin{equation}
    \theta_x=\displaystyle\frac{f_t \lambda}{V_x} \;\;{\rm and}\;\; \theta_y=\pm\sqrt{\displaystyle\frac{2cf_\nu}{Z}-\left(\frac{f_t \lambda}{V_x}\right)^2}
\end{equation}
Here we assume the velocity is in the x direction. The Jacobian of the transformation is J$=$1/$\theta_y$ which has a half-order singularity where $\theta_y=0$. This singularity is the sharp arc. We have simplified the equations without loss of generality by taking the case of the scattering medium a distance Z from the observer and an infinitely distant source. We can demonstrate this mapping for a Kolmogorov brightness distribution as shown in Extended Data\ref{fig:sim_ss}. Here we have plotted three cases left to right: isotropic; anisotropic aligned with V; and anisotropic perpendicular to V. In these simulations, the anisotropic cases have an axial ratio $A_R = 3$. We have included a phase gradient typical of that in seen in the observed secondary spectra in each simulation. In anisotropic scattering cases, the shape of the primary arc is independent of the anisotropy, but the power distribution inside the arc depends strongly on it. From Extended Data\ref{fig:sim_ss}, an axial ratio for the scattering structures of just 3, with the major axis of scattering irregularities (i.e. the minor axis of the brightness distribution) aligned with the velocity, results in a central ridge that matches the observations well. Neither the isotropic case nor a perpendicular anisotropy result in a central ridge. For deviations in alignment greater than about 10$^\circ$, the ridge would be noticeably displaced from the central axis.

Based on the observed value of scintillation timescale at 1400~MHz, we can derive the spatial scale and the rms scattering angle in the direction parallel to the velocity:

\begin{equation}
   \theta_{\rm par}=\displaystyle\frac{c}{2\pi \nu_c \Delta \tau_d [V_{\rm psr, \perp} (1-s)-V_{\rm scr, \perp}]}=0.70\pm0.10~{\rm mas}. 
\end{equation}
The rms scattering angle in the perpendicular direction will be $\theta_{\rm perp} = A_R\ \theta_{\rm par}$, where $A_R$ is the axial ratio of an asymmetric scattering structure. Furthermore, we can use the measured bandwidth to estimate $A_R$ as follows

\begin{equation}
    \Delta\nu_d = \displaystyle\frac{1}{2 \pi \tau_{\rm sc}} =\displaystyle \frac{c}{\pi D_{\rm ps}(\theta_{\rm par}^2 + \theta_{\rm perp}^2)}
\end{equation}
where $\tau_{\rm sc}$ is the profile scatter-broadening timescale. Hence

\begin{equation}
    A_R^2  = \displaystyle\frac{c}{\pi \Delta\nu_d \, D_{\rm ps}\, \theta_{\rm par}^2}-1
\end{equation}
Substituting for $\Delta\nu_d\sim0.86$~MHz and $\theta_{\rm par}\sim0.70~{\rm mas}$ at 1400~MHz, we obtain $A_R\sim3$ which matches well with the simulations above.

\end{methods}

%% Here is the endmatter stuff: Supplementary Info, etc.
%% Use \item's to separate, default label is "Acknowledgements"

\begin{addendum}
\item[Correspondence and requests for materials] This should be addressed to Jumei Yao (yaojumei@xao.ac.cn).  \item[Acknowledgements]This work was supported by the National Natural Science Foundation of China Grant No.11988101, 11903049, 12041304, 12041303, 11873067, 11690024, U1831104, the CAS-MPG LEGACY project and the Strategic Priority Research Program of the Chinese Academy of Sciences Grant No. XDB23000000 and the National SKA Program of China No. 2020SKA0120200. JMY acknowledges support from the CAS “Light of West China” Program 2017-XBQNXZ-B-022 and the Tianchi Doctoral Program 2017. WWZ was supported by the Chinese Academy of Science Pioneer Hundred Talents Program. We thank the anonymous reviewer for helpful suggestions, and Hans-Thomas Janka and Shi Dai for valuable comments on earlier versions of the manuscript.
 \item[Author contributions] JMY led the project. WWZ, DL, NW, JGL, and DRS were involved in discussing the proposal for these observations. PW and CCM helped with data analysis and MY made the two 3D plots. RNM, WWZ, WAC and DL had major contributions to the preparation of the manuscript. DRS commented on ISS data analysis, and YF, WMY and MK helped with polarization calibrations. All authors reviewed, discussed, and commented on the present results and the manuscript.  
 \item[Additional information] Supplementary information is available in the online version of the paper.
 \item[Competing financial interests] The authors declare that they have no competing financial interests.  
 \end{addendum}

%%
%% TABLES
%%
%% If there are any tables, put them here.
%%
\newpage
\appendix

\renewcommand{\thefigure}{ Figure \arabic{figure}}   
\renewcommand{\thetable}{Table \arabic{table}} 
\renewcommand{\figurename}{Extended Data}
\renewcommand{\tablename}{Extended Data}

\begin{figure*}
\centering
\includegraphics[width=10 cm, angle=270]{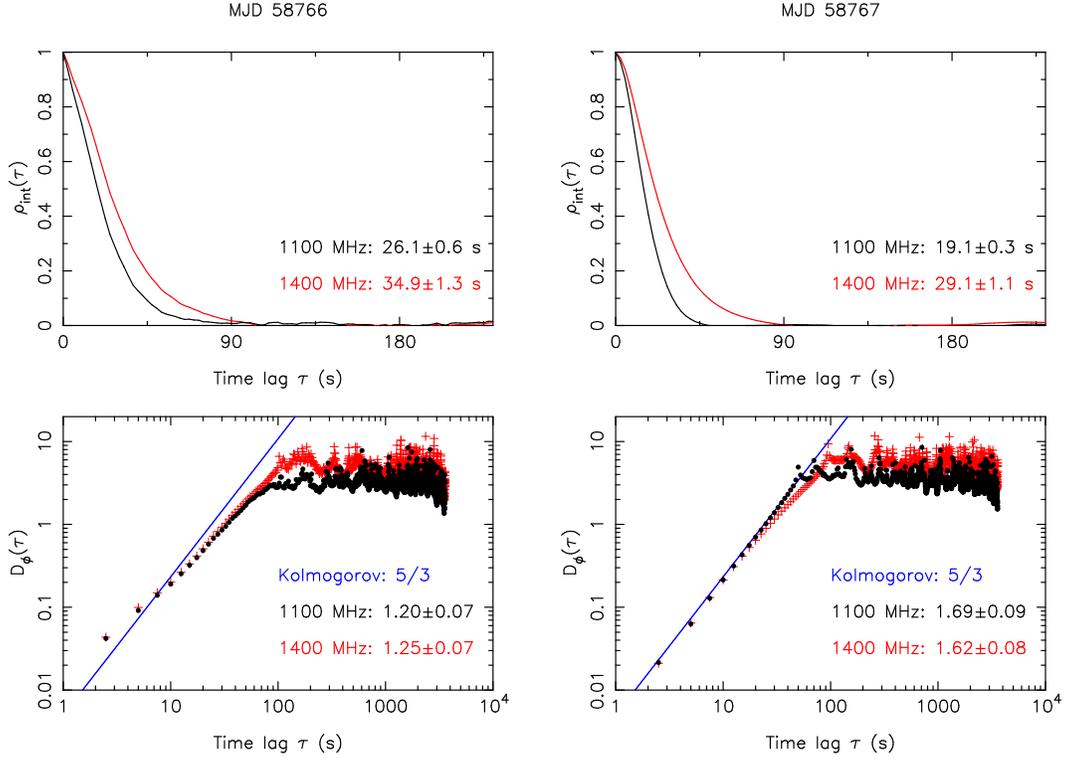}
\caption{One-dimensional time-domain ACFs for dynamic spectra centred at 1100 and 1400~MHz (upper panel) and the corresponding structure functions (lower panel) for MJDs 58766 and 58767. To facilitate comparison of the linear parts of the structure functions for the two bands, the $D_{\phi}$ axis for the 1100~MHz structure function has been scaled by (1100/1400)$^2$.}
\label{fig:acf_sf}
\end{figure*}

\begin{figure*}
\centering
\includegraphics[width=6 cm, angle=270]{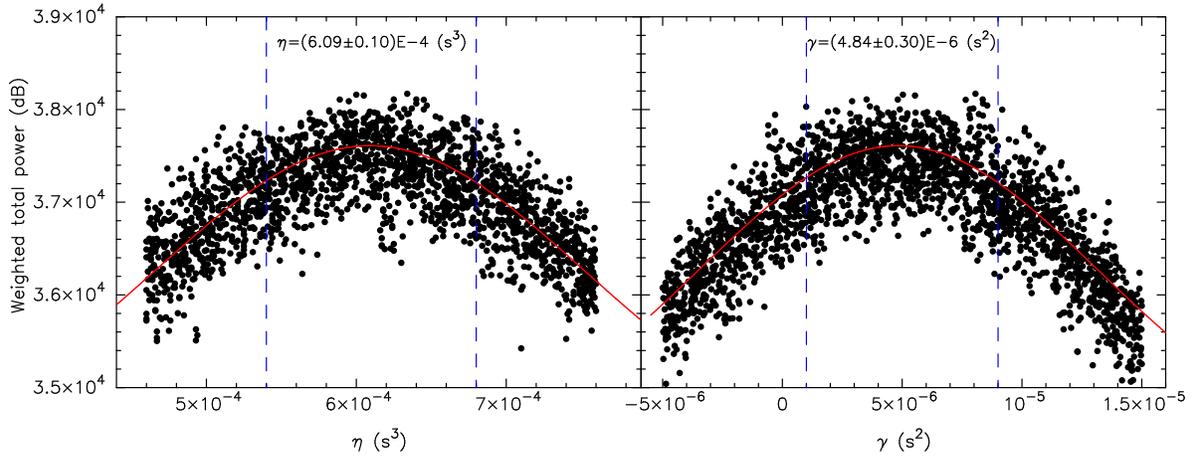}
\caption{Distribution of the summed arc power in the $\eta$ direction (left panel) and the $\gamma$ direction (right panel). The vertical dashed lines delineate the $\eta$ and $\gamma$ ranges used for the two-dimensional fit to the power and the plotted points are from within these ranges.  The red lines are cuts through the 2D Gaussian fit at the best-fit value of the other coordinate.}
\label{fig:arc_fit}
\end{figure*}

\begin{figure*}
\centering
\includegraphics[width=120mm]{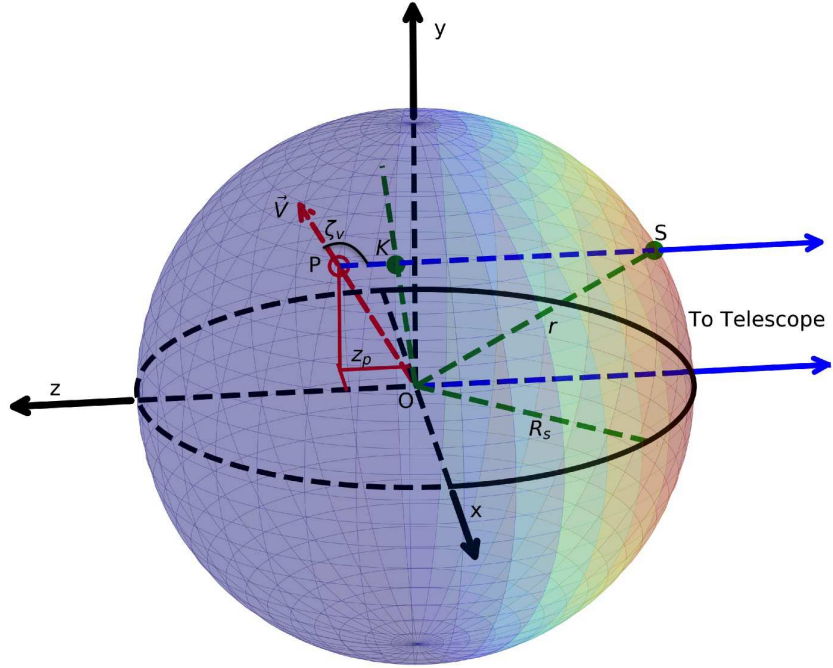}
\caption{The adopted coordinate system for the SNR S147 -- PSR J0538+2817 system. The coordinate system is centred at the geometric centre of SNR S147 (O) with the +x axis in the direction of increasing right ascension (E), the +y axis toward north, and the +z axis away from the observer. We mark the current pulsar position as $P$ ($x_p$, $y_p$, $z_p$) and its projection on the x-y plane as $K$ ($x_p$, $y_p$, 0), and the scattering screen as $S$ ($x_s$, $y_s$, $z_s$). The inclination of the pulsar velocity $\vec{V}$ to the line of sight is $\zeta_v$, $r$ is the distance of the scattering screen from the origin and the SNR shell radius is $R_s$.}
\label{fig:coord}
\end{figure*}

\begin{figure*}
\centering
\includegraphics[width=160mm]{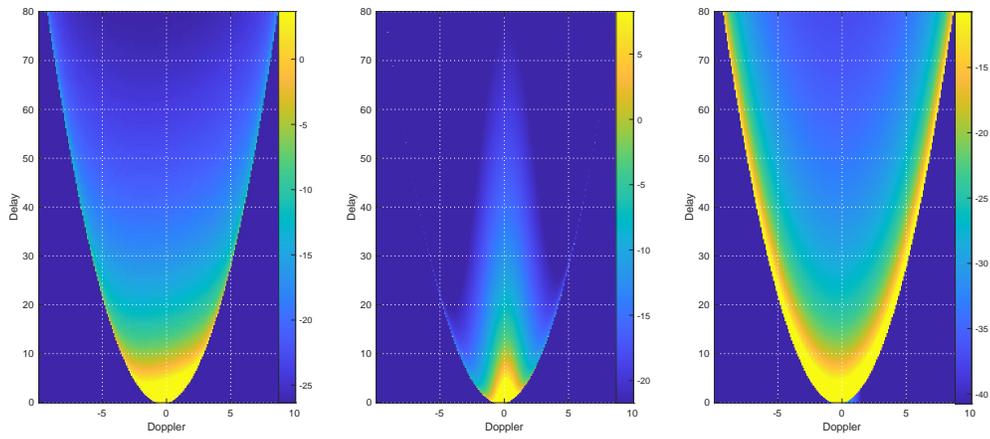}
\caption{Simulated secondary spectra: isotropic (left); anisotropic aligned with V, as proposed for the observations (centre); anisotropic perpendicular to V (right), with a logarithmic (dB) colour scale. The axial ratio of the anisotropic cases is 3. Each has the same phase gradient, set equal to the value of $\gamma$ inferred from the observations.}
\label{fig:sim_ss}
\end{figure*}

\begin{table*}
\tiny
\caption{The value of $\eta$ and $\gamma$ and seven parameters derived from 1400~MHz observations at four observational epochs.}\label{tb:obs_epochs}
\centering
\begin{tabular}{ccccccccccc}
	\hline
	MJD & Time &$\eta$ & $\gamma$ & s &$D_{\rm ps}$&$z_p$  & $\zeta_V$& $V_z$ & $V_{\rm tot}$ & $\theta_g$ \\
    & (min)  &  (s$^3$)  & (s$^2$) &    & (pc)& (pc) & (deg)& (km\,s$^{-1}$) & (km\,s$^{-1}$)& ($\mu$as)\\
    \hline
58635 & 45 & (6.24$\pm$0.14)$\times 10^{-4}$& ($-$3.24$\pm$0.67)$\times 10^{-6}$& $0.0247^{+0.0043}_{-0.0043}$ &$32.9^{+7.4}_{-7.4}$&  $3.6^{+5.9}_{-5.6}$ &$115^{+15}_{-30}$  &  $101^{+166}_{-157}$& 408$^{+83}_{-57}$& $-8.0^{+1.7}_{-1.7}$\\
58733 & 120 & (6.08$\pm$0.06)$\times 10^{-4}$& (3.46$\pm$0.28)$\times 10^{-6}$& $0.0241^{+0.0040}_{-0.0040}$ &$32.1^{+7.0}_{-7.0}$&  $2.9^{+5.5}_{-5.2}$ &$110^{+16}_{-28}$  &  $81^{+154}_{-147}$& 406$^{+78}_{-57}$& $8.7^{+0.8}_{-0.8}$\\ 
58766 & 60 &  (6.01$\pm$0.05)$\times 10^{-4}$& (1.61$\pm$0.24)$\times 10^{-6}$&$0.0238^{+0.0040}_{-0.0040}$ &$31.7^{+7.0}_{-7.0}$&  $2.5^{+5.5}_{-5.2}$ &$108^{+17}_{-28}$  &  $71^{+154}_{-147}$& 406$^{+77}_{-58}$& $4.0^{+0.6}_{-0.6}$\\
58767 & 60 &  (6.09$\pm$0.10)$\times 10^{-4}$& (4.84$\pm$0.30)$\times 10^{-6}$& $0.0241^{+0.0041}_{-0.0041}$ &$32.1^{+7.1}_{-7.1}$&  $2.9^{+5.6}_{-5.4}$ &$110^{+16}_{-29}$  &  $81^{+158}_{-150}$& 407$^{+79}_{-57}$& $12.1^{+0.8}_{-0.8}$\\
	\hline
	\end{tabular}
\end{table*}

\end{document}